\documentclass[preprint]{aastex}

\shorttitle{Close binary stars.~XIV}
\shortauthors{Pribulla \& et al.}

\begin{document}

\title{Radial Velocity Studies of Close Binary
Stars.~XIV\footnote{Based on the data obtained at the David Dunlap
Observatory, University of Toronto.}}

\author{}
\author{
Theodor Pribulla\altaffilmark{1},
Slavek M. Rucinski, Heide DeBond, Archie de Ridder, Toomas Karmo,\\ 
J.R. Thomson., B. Croll}
\affil{David Dunlap Observatory, University of Toronto \\
P.O.~Box 360, Richmond Hill, Ontario, Canada L4C~4Y6}
\email{pribulla@ta3.sk,(rucinski,debond,ridder,karmo,croll)@astro.utoronto.ca}
\author{Waldemar Og{\l}oza}
\affil{Mt. Suhora Observatory of the Pedagogical University\\
ul.~Podchora\.{z}ych 2, 30--084 Krak\'ow, Poland}
\email{ogloza@ap.krakow.pl}
\author{Bogumil Pilecki}
\affil{Warsaw University Astronomical Observatory, 
 Al. Ujazdowskie 4, 00-478 Warszawa, Poland}
\email{pilecki@astrouw.edu.pl}
\author{Michal Siwak}
\affil{Astronomical Observatory, Jagiellonian University, ul. Orla 171,
       30--244 Krak\'ow, Poland}
\email{siwak@oa.uj.edu.pl}

\altaffiltext{2}
{Astronomical Institute, Slovak Academy of Sciences, 
059 60 Tatransk\'a Lomnica, Slovakia}

\begin{abstract}
Radial-velocity measurements and sine-curve fits to the orbital radial
velocity variations are presented for ten close binary systems: TZ~Boo,
VW~Boo, EL~Boo, VZ~CVn, GK~Cep, RW~Com, V2610~Oph, V1387~Ori, AU~Ser, and
FT~UMa. Our spectroscopy revealed two quadruple systems, 
TZ~Boo and V2610~Oph, while three stars showing small 
photometric amplitudes, EL~Boo, V1387~Ori, and
FT~UMa, were found to be triple systems. 
GK~Cep is close binary with a faint third component.

While most of the studied eclipsing systems are contact binaries, VZ~CVn 
and GK~Cep are detached or semi-detached double-lined binaries, and 
EL~Boo, V1387~Ori and FT~UMa are close binaries of uncertain binary type. 
The large fraction of triple and
quadruple systems found in this sample supports the hypothesis of formation
of close binaries in multiple stellar systems; it also demonstrates that
low photometric amplitude binaries are a fertile ground for further
discoveries of multiple systems.
\end{abstract}


\section{INTRODUCTION}
\label{sec1}

This paper is a continuation of a series of papers (Papers
I -- XIII) of radial-velocity studies of close binary stars
and presents data for the thirteenth group of 
ten close binary stars observed at the David Dunlap 
Observatory (DDO). Because of the closure of the
observatory, it is most likely the last paper which
retains the usual format of 10 new orbits per
paper; the last
paper, Paper XV, will conclude the series
with results for partly covered systems and for variable
stars which were found not to be binaries. For full references to the
previous papers, see the last paper by \citet[Paper XIII]{ddo13};
for technical details and conventions, for preliminary
estimates of uncertainties, and for a description of the
broadening functions (BFs) technique, see the interim summary
paper \citet[ Paper VII]{ddo7}. The DDO studies have
used the efficient program of \citet{pych2004}
for removal of cosmic rays from 2-D images.

All data used in the present paper were obtained using the broadening
functions extracted from the region of the Mg~I triplet at
5184~\AA, as in most of the previous papers, using the new 
2160 lines/mm grating acquired at DDO in August 2005. The radial-velocity 
(hereafter RV) observations reported in this paper have been collected 
between May 2006 and the memorable day of
July 2, 2008 when the David Dunlap Observatory ceased to operate. 
The ranges of dates for individual systems can 
be found in Table~\ref{tab1}; for TZ~Boo, where we 
used the smoothed BFs, the dates are in Table~\ref{tab3} giving
RVs for the companion of the short-period binary.

Throughout our program, selection of the targets was quasi-random: 
At a given time, we observed a few dozen close binary 
systems with periods usually shorter than one day, 
brighter than 10 -- 11 magnitude and with
declinations $>-20^\circ$; we published the results in 
groups of ten systems
as soon as reasonable orbital elements were 
obtained from measurements evenly
distributed in orbital phases. 

Among the present targets, three systems, AU~Ser, VW~Boo and VZ~CVn, have 
had reliable radial velocity orbits previously published. In addition,
EL~Boo and FT~UMa were originally classified as pulsating 
variables while V1387~Ori does not have any ground-based 
photometry, but only Hipparcos data. More details are given
in Section~\ref{sec2} in the descriptions of the individual
stars.

The RVs for the short period binaries reported in this paper 
were determined by fitting the double rotational profiles to the
extracted BFs, as explained in \citet{ddo11}. 
Similarly as in our previous papers dealing with multiple systems
(here the six cases of TZ~Boo, EL~Boo, GK~Cep, V2610~Oph, V1387~Ori, 
and FT~UMa), the RVs for the eclipsing pair were obtained 
after removal of the slowly rotating components, 
as was described most recently in \citet{ddo13}. In the case of GK~Cep, 
where the third component is faint and not 
well visible in all spectra, its average
profile was subtracted before the binary star analysis.

As in other papers of this series, whenever possible, we estimated spectral
types of the program stars using new classification spectra centered
at 4200 or 4400 \AA. Good quality, multiple 
classification spectra were obtained 
before the closure of the observatory only
for EL~Boo, VZ~CVn, GK~Cep, V1387~Ori, and FT~UMa; for the
remaining targets the classification spectra were obtained only once and
in poorer conditions so that the spectral types are
less reliable. The spectral types were compared 
with the mean $(B-V)$ color indices usually taken from the Tycho-2 catalog 
\citep{Tycho2} and the photometric estimates of the spectral types using the 
relations of \citet{Bessell1979}. In this paper we also made use of infrared 
colors determined from the 2$\mu$ All Sky Survey (2MASS, \citealt{2MASS}). 
Especially 
useful is the $J-K$ color index, which is 
monotonically rising from the early 
spectral types to about M0V \citep{cox2000}. This infrared color is 
less affected by the interstellar absorption than the $B-V$ index.
Parallaxes cited throughout the paper were adopted from the new reduction
of the Hipparcos raw data \citep{newhipp} which supersedes the original
reductions \citep{hip}.

This paper is structured in a way similar to that 
of previous papers, in that
most of the data for the observed binaries are in 
two tables consisting of the
RV measurements in Table~\ref{tab1} and 
of their preliminary sine-curve solutions
in Table~\ref{tab2}. Radial velocities and the 
corresponding spectroscopic orbits 
for all ten systems are shown in phase diagrams in 
Figures~\ref{fig1} -- \ref{fig3}.
The measured RV's are listed in Table~\ref{tab1}. 
Table~\ref{tab2} contains also our new spectral 
classifications of the program 
objects. Section~\ref{sec2} of the paper contains 
summaries of previous studies 
for individual systems and comments on the new data. 
Examples of BFs of individual 
systems extracted from spectra observed close to quadratures are shown in 
Fig.~\ref{fig4}.

The data in Table~\ref{tab2} are organized in the same manner as in
the previous papers of this series. 
In addition to the parameters of spectroscopic orbits,
the table provides information about the relation between the
spectroscopically observed upper conjunction of the more massive
component, $T_0$ (not necessarily the primary eclipse)
and the recent photometric determinations of the primary
minimum in the form of the $O-C$ deviations for the number of
elapsed periods $E$. The reference ephemerides were taken from
various sources: for EL~Boo, and FT~UMa, we doubled the Hipparcos 
period and shifted the instant of the maximum by $0.25P$; for V1387~Ori 
we used the Hipparcos ephemeris; and for V2610~Oph, 
it has been adopted from \citet{wils2003}. 
For the rest of the systems, the ephemerides given in the on-line version 
of ``An Atlas of O-C diagrams of eclipsing binary 
stars''\footnote{http://www.as.wsp.krakow.pl/ephem/} \citep{Kreiner2004}
were adopted. As the on-line ephemerides are frequently updated, we
give those used for the computation of the $O-C$ residuals below
Table~\ref{tab2} (up-to-date as of July 2008). The deeper eclipse in W-type
contact binary systems corresponds to the lower conjunction of the more 
massive component; in such cases the epoch in Table~\ref{tab2} is a 
half-integer number.

\section{RESULTS FOR INDIVIDUAL SYSTEMS}
\label{sec2}

\subsection{TZ~Boo}

TZ~Boo (HIP~74061, BD+40\degr 2857) is a well known contact binary
discovered by \citet{guth1926}. The system is unusual:
Its depths of the minima change, switching between the
A and W-type light curve (LC) types and the LC shows very large 
variations of its shape \citep{hoff1978a}. 
In spite of total eclipses, the large 
variability of the LC has made a reliable 
photometric solution of the system
impossible; thus, there exists no modern LC synthesis solution 
assuming the Roche model. The
orbital period of the system is also variable. The continuous orbital period 
decrease of TZ~Boo, accompanied by wave-like variations was interpreted
by \citet{priruc2006} in terms of light-time effect caused by a third 
component on a 34 years orbit. But an adaptive 
optics search for close visual
companions to contact binaries \citep{ruc2007} did not show any close
companion to TZ~Boo.

TZ~Boo is a very difficult system for spectroscopic studies.
\citet{chang1948} found that spectral lines do not double in quadratures,
but noticed that the measured RV of the system varies in the range of
about 60 km~s$^{-1}$. The author did not find any correlation between
the RV and the predicted photometric orbital phase. Later
\citet{hoff1978b} tried to interpret the results of \citet{chang1948} by
systemic velocity changes induced by a third body on a 10.31 day orbit.
\citet{hoff1978b} added a new set of spectroscopic observations
to the older observations, but could
not solve the problem. The author estimated the mass ratio as $q = 0.2$ 
(which in fact is identical to our current result) and
interpreted the inexplicable behavior of the profiles by the 
enhanced activity in the system. 
Finally, \cite{lean1983} obtained the first, 
seemingly fully satisfactory
spectroscopic orbit of the system: $V_0 = -36.7$ km~s$^{-1}$, 
$K_1 = 33 \pm 7$ km~s$^{-1}$, and $K_2 = 249 \pm 38$ km~s$^{-1}$ 
resulting in $q = 0.13 \pm 0.03$. The authors did not notice a third 
component in the profile that -- as we suspect now --
was blended with the primary component; 
unfortunately this means that this 
spectroscopic orbital solution is of limited use.

Our spectroscopy shows that TZ~Boo is either 
a triple system with a third body
on a 9.48 day orbit or -- more likely -- a 
quadruple system consisting of
a contact eclipsing binary and a second, 
non-eclipsing, single-line binary with
the period of 9.48 days. The system was 
found to be fairly difficult for the DDO
telescope due to the short orbital period of the contact binary, 
$P_{12} = 0.297$ days, its relatively low brightness, 
$V_{max} = 10.4$, and the obviously composite spectra.
To determine reliable parameters, we took as many as 215 spectra spanning
more than two years. First, we fitted a three-Gaussian model to each BF and 
subtracted the third component contribution. The BFs of the contact binary 
were noisy so we decided to smooth and rebin them with a 0.01 step
in orbital phase. 
The resulting mass ratio, $q=0.207 \pm 0.005$, is rather small and
inconsistent with the large photometric amplitude of the 
system (note however the disparity of the values: 0.59 mag in the 
General Catalogue of Variable Stars, 0.39 mag in Hipparcos photometry, 
and 0.35 mag according to Fig.~1 of \citealt{schaub1990}). 
It is possible that the pronounced LC changes and the W/A-type
switching are caused by the activity and/or by relatively slow (and of
unexplained nature) pulsations of the third component.
Also, the possibility of
eclipses in the second binary cannot be ruled out. The changes of the LC
amplitude -- as seen best in Fig.~2 of \citet{awad2006} -- 
could also be caused
by significant long-term brightness changes of the third/fourth component. 
Another explanation could be the precession of the eclipsing-pair
orbit, but this is less likely because all the 
observed LCs have shown total eclipses.
The system deserves a dedicated study of the light curve changes
in a standardized photometric system to relate the relative LC shape
changes to its overall brightness.

The orbit of the second (or outer) binary is circular -- which is
rather surprising considering the long orbital period --
and is very well defined with $K_3 = 43.15 \pm 0.14$ km~s$^{-1}$ 
(see Fig.~\ref{fig5} and Table~\ref{tab4}). 
The center-of-mass velocities of the contact binary, 
$V_0^{12} = -46.57 \pm 0.90$ km~s$^{-1}$, and the non-eclipsing binary, 
$V_0^{34} = -54.64 \pm 0.12$ km~s$^{-1}$ are different, indicating 
a slow orbital motion. The light contribution of the third 
component around the 
quadratures of the contact binary is $L_3/(L_1 + L_2) = 0.28 \pm 0.05$. 
Because of the well recognized problems with a
proper continuum rectification of the spectra of combined
broad and sharp line components
\citep{rucpri2008}, we regard 
this estimate of the third light as an upper limit. 

The RVs of the contact pair show a rather large point-to-point scatter
of about 10 km~s$^{-1}$. The scatter does not decrease under the
assumption that TZ~Boo is a tight triple system and that 
the contact binary revolves around the common center of 
mass with the third component.
Although with the present data we cannot fully exclude this possibility,
it is much more probable that TZ~Boo is a hierarchic quadruple 
consisting of two binaries revolving
in 34-year orbit, as indicated by the cyclic period changes. 

2MASS infrared color, $J-K=0.454$ corresponds to G7V spectral type.
Our single spectral classification spectrum indicates a range 
of admissible spectral types between late F and G5.
The Hipparcos parallax, $\pi = 6.63 \pm 1.54$ mas
is unfortunately too uncertain to be of value.

\subsection{VW~Boo}

VW~Boo (HIP~69826, GSC~908~1170) is a short-period 
($P = 0.3422$ days) close binary
that was discovered by \citet{hoff1935}. 
\citet{binn1973} obtained and analyzed
$BV$ photoelectric photometry of the system, 
which shows that components have 
rather different temperatures for such a close binary. 
Therefore, VW~Boo belongs 
to a small group of short-period contact binaries that have poor thermal 
contact (others are e.g., CN~And, FT~Lup, V432~Per or AU~Ser).
 
\citet{rain1990} re-analyzed LCs of \citet{binn1973} and obtained the
first spectroscopic data for the system. 
The authors interpreted the LCs by a contact configuration with a 
hot spot on the secondary component, about 640~K hotter than the surrounding
photosphere. The CCF analysis of the spectra yielded 
the following spectroscopic 
elements for VW~Boo: $K_1 = 99.2 \pm 2.1$ km~s$^{-1}$, 
$K_2 = 230.1 \pm 5.4$ 
km~s$^{-1}$ (resulting in $q = 0.428 \pm 0.030$), 
$V_{01} = 21.5 \pm 1.5$ km~s$^{-1}$
and $V_{02} = 26.3 \pm 4.2$ km~s$^{-1}$ 
(determined separately for the two components).
Independent observations of \citet{hriv1993} 
resulted in $q = 0.45$ and the total
projected mass of $(M_1 + M_2) \sin^3 i = 1.34$ M$_\odot$.

Our spectroscopic elements (Table~\ref{tab2}) 
are in a good accord with the results of \citet{rain1990}.
However, the extracted BFs do not show any irregularities which could
indicate the presence of either hot or cool spots on the surface. 
The rotational profiles of the components in BFs 
taken at the orbital quadratures suggest 
that system is either in marginal physical contact
or is a detached one. 

Although the system was observed by Hipparcos, its trigonometric parallax, 
$\pi = 2.12 \pm 2.52$ mas is too uncertain to be of any use. 
The 2MASS infrared color of the system, $J-K = 0.467$ corresponds 
to the G8V spectral type which agrees with our classification
G5V.

\subsection{EL~Boo}

Variability of EL~Boo (HIP~72391, SAO~101223) 
was detected during the Hipparcos mission, 
where it was classified as a $\delta$~Sct 
variable with $P$ = 0.2068860 days. Later, 
it was identified as an X-ray source \citep{mick2006}. 
Observations of only two minima 
are available in the literature \citep{sena2007,hubs2007}. 
In the ASAS 
database\footnote{http://www.astrouw.edu.pl/asas/} 
it is classified as an eclipsing binary. 
The ASAS light-curve indicates a W~UMa-type 
classification and total eclipses. 
No other photometry or spectroscopy of the 
target has been published yet.

The BFs (see Fig.~\ref{fig4}) clearly show 
that EL~Boo is a triple system containing  
a very close (possibly contact) binary and a slowly-rotating third star. 
The system was not previously known to be a visual binary.
The orbital period of the eclipsing pair, 
$P = 0.413772$, is two times longer than the previous 
Hipparcos (pulsation) period. The third component 
contributes about a half of light to the total brightness of the system,
$L_3/(L_1 + L_2) = 1.00 \pm 0.08$. The RV of the 
third component was found to be constant, 
$RV_3 = -22.6 \pm 1.1$ km~s$^{-1}$,
and rather close to the systemic velocity of the close eclipsing binary, 
$V_0 = -24.6 \pm 1.1$ km~s$^{-1}$ indicating a physical bond. 
The projected rotation  velocity of the third component 
is low but detectable even at our rather moderate 
spectral resolution, $v \sin i \approx$ 23 km~s$^{-1}$.
The outer orbital period of this triple 
must be at least several years or longer. 
 
The relatively small photometric amplitude of the system, 0.19 mag, 
results from the significant light contribution of the third component 
and from the rather low mass ratio of the binary, $q = 0.248 \pm 0.007$.
On the other hand, the projected mass of the close binary, 
$(M_1 + M_2) \sin^3 i = 1.448 \pm 0.028$ M$_\odot$ is high for the 
F5V spectral type indicating a high inclination angle and the implied
possibility of total eclipses. The photometric amplitude
of the close binary, without the diluting effect of the third light, is
expected to be about 0.42 mag, which is unexpectedly large, 
even if eclipses of the close binary were total; this discrepancy, 
which is similar to that in TZ~Boo, indicates that 
we may have overestimated the amount 
of the third light. The system definitely 
deserves new, high-precision photometric observations. 

The published F8 spectral type is fairly late for a contact binary 
with a 0.414 day orbital period; however this mainly reflects 
the contribution of the third component because 
spectral lines of the contact binary
are highly broadened by the fast rotation and are practically 
invisible in classification dispersion
spectra. The 2MASS infrared color, $J-K = 0.282$, corresponds to the 
F5V spectral type and our own classification spectra also
indicate the F5V type. The Hipparcos parallax of the system,
$5.16 \pm 1.69$ mas is too small and uncertain to draw any conclusions.

\subsection{VZ~CVn}

VZ~CVn (HD~117777, HIP~66017) is a rather bright ($V_{max}$ = 9.35),
detached eclipsing binary. It was discovered to be 
variable by \citet{stro1960}. The system is known to show a
slow intrinsic variability in 
the LC (see \citealt{popp1988}). Recently, \citet{iban2007} --
after removing the eclipse variability --
detected $\gamma$~Dor-type oscillations of the primary 
component and determined the first set of photometric elements. 
As this star has an orbital period of 0.842 days, it is difficult 
to cover all orbital phases for this target from one site; 
in addition 
intrinsic night-to-night variations 
of a few hundreds of a magnitude make a consistent 
LC solution almost impossible using ground-based photometry. 
The system is therefore a prime candidate for satellite,
continuous photometric observations.

VZ~CVn was observed spectroscopically by 
\citet{popp1988}, who determined the first
radial velocity orbit from photographic spectra: 
$K_1 = 144.1 \pm 1.2$ km~s$^{-1}$, 
$K_2 = 185.4 \pm 3.0$ km~s$^{-1}$ and the separate systemic velocities 
of the two components as 
$V_{01} = -20.7 \pm 0.7$ km~s$^{-1}$, and 
$V_{02} = -21.7 \pm 2.0$ km~s$^{-1}$.

Our observations provide very well defined BFs leading to spectroscopic
elements that are mostly consistent with those of \citet{popp1988}. 
The largest discrepancy is the smaller value of the semi-amplitude
$K_2$. The spectroscopic orbit is, however, of 
excellent quality, leading to a 
total projected mass of $(M_1 + M_2) \sin^3 i = 2.676 \pm 0.007 M_\odot$, 
which corresponds to the relative precision as good as 0.26~\%. 
The broadening functions are well separated in 
orbital quadratures, thus supporting the 
photometric classification of the system as a detached binary.

The 2MASS infrared color of the system, $J-K = 0.216$ 
indicates the F2V spectral type.
Our classification spectra correspond to F0V. 
The Hipparcos parallax is small, $2.23 \pm 1.22$ mas and
too inaccurate to be of use.

\subsection{GK~Cep}

The bright variable star GK~Cep 
(HD~205372, HIP~106226; $V_{max} = 6.99$), 
originally classified as an RR~Lyr pulsator, 
was later found to be an eclipsing binary with
a 0.483 days period (see \citealt{stro1963}). 
\citet{bart1965} observed the system 
photometrically and spectroscopically and determined 
the correct orbital period, 
$P$ = 0.936171 days, and the
first set of spectroscopic elements: $V_0 = -22$ km~s$^{-1}$,
$K_1 = 172$ km~s$^{-1}$, $K_2 = 187$ km~s$^{-1}$. 
The authors classified GK~Cep
as a $\beta$~Lyrae variable and, with the
photometrically determined inclination angle,
$i$ = 71\degr, they determined the masses of 
the components, $M_1$ = 2.7 $M_\odot$
and $M_1$ = 2.5 $M_\odot$. Later, \citet{hutch1973} 
analyzed the LCs of the system
assuming the Roche model and found that the system is a close, 
but detached binary; the more massive component is the cooler one.

The orbital period analysis of \citet{krein1990} 
suggests the presence of an
invisible third body in the system on a 18.8 year orbit. 
Assuming that the orbits are co-planar,
the authors estimated the mass of the third component as 1.34 M$_\odot$.

Our spectroscopy confirms the existence 
of the third component. It is, however,
faint with its light contribution as small 
as $L_3/(L_1 + L_2) \approx 0.025$. 
The profile of the third component is not well 
visible in all of the BFs because it 
blends with the more massive component, especially 
around the second quadrature. 
This circumstance complicated our usual approach, i.e.\ the fitting 
of a triple Gaussian model to 
the observed BFs. Hence we averaged all BFs (in the 
heliocentric RV system) 
and the mean BF was used to define 
the third component contribution which was subsequently  
subtracted from all individual BFs. 
The RV of the third component, $RV_3 = -19.3$ km~s$^{-1}$,
is rather different from the systemic
velocity of the system, $V_0 = -31.06 \pm 0.45$ km~s$^{-1}$,
indicating that the faint companion is not physically
related to the binary.

Our spectroscopic elements (Table~\ref{tab2}), 
largely agree with those of 
\citet{bart1965}. Our RV semi-amplitudes are, 
however, smaller resulting in an estimate of the 
total mass of the system that is smaller by about 15\% 
than that of \citet{bart1965}. The very large mass ratio, 
$q = 0.913 \pm 0.005$, which would be unusual if it were a contact binary, 
supports the result of \citet{hutch1973} that the system is detached.

The Hipparcos parallax $\pi = 5.17 \pm 0.32$ mas 
(the distance $d = 193 \pm 11$ pc) is precise thanks 
to the high ecliptical latitude. The 2MASS 
infrared color, $J-K = 0.007$ corresponds to the A1V spectral type, 
which is in good accord with the previous spectral classification, 
A2V (Hill et al., 1975). Our classification 
spectra give a slightly earlier spectral type, A0V. If the components 
of GK~Cep are normal main-sequence stars, the best accord with the 
Hipparcos parallax is achieved for the A2V spectral type.

\subsection{RW~Com}

RW~Com (HIP~61243, $V_{max} \approx 11.0$ mag, sp.\ type G5 -- G8) is one
of the shortest-period ($P = 0.237346$ day) W~UMa systems. 
Its variability was first noticed by \citet{jord1923}.
This late-type contact binary is known to show an
enhanced surface activity resulting in an asymmetric 
LC \citep{milo1987}.
The orbital period of the system shows a continuous decrease of
$dP/dt = -6.06 \times 10^{-8}$ day/year \citep{qian2002} 
accompanied by a sinusoidal variation with a 16-year periodicity. 

The first spectroscopy of RW~Com was carried 
out by \citet{stru1950} who found 
the H \& K Ca~II lines in emission. 
Later, \citet{milo1985} presented the first RV orbit
of the system: $K_1 + K_2 = 304 \pm 5$ km~s$^{-1}$, 
$(M_1 + M_2) \sin^3 i = 0.69 \pm 0.04 M_\odot$, $M_1/M_2 = 2.9 \pm 0.2$ 
(this corresponds to $q = 0.349 \pm 0.022$) 
and $V_0 = -53 \pm 4$ km~s$^{-1}$. Unfortunately, 
the spectra were of a poor quality, with relatively long 
exposure times typically lasting 20 minutes, 
which resulted in heavy blending of the cross-correlation functions 
used for radial velocity determinations. 

Because of the short orbital period of the system, 
we kept all our exposures
equal to 500 sec (2.43\% of the orbital period). Our data, however, by far 
supersede those presented by \citet{milo1985}. 
The resulting parameters given in Table~\ref{tab2} are inconsistent with 
those given by \citet{milo1985}; 
the most striking difference is our much higher 
mass ratio, $q = 0.471 \pm 0.006$, and the substantially larger total
projected mass, $(M_1 + M_2) \sin^3 i = 1.052 \pm 0.013 M_\odot$. 
It is interesting to note, that very
similar mass ratios, all close to 0.5, 
have been found for two other contact, very late-type binaries with 
extremely short orbital periods: CC~Com ($q = 0.527$, \citealt{ddo12})
and GSC~1387~475 ($q = 0.474$, \citealt{rucpri2008}).
The broadening functions 
do not show any trace of the third component which was 
indicated by the cyclic period change. 
This sets the upper limit on its luminosity at 
about $L_3/(L_1 + L_2) < 0.03$.

Using the Hipparcos parallax, 
$\pi = 14.33 \pm 3.36$ mas, and the known maximum 
apparent magnitude, $V_{max} = 11.0$,
one obtains $M_V = 6.8 \pm 0.8$. 
With the calibration of \citet{rd1997} and using
$B-V=1.07$ from the TYCHO2 catalogue, one obtains $M_V^{cal} = 6.12$. The 
2MASS infrared color of RW~Com, $J-K = 0.618$ 
corresponds to the K2V spectral type,
which is inconsistent with the G5--G8 spectral type 
estimated by \citet{milo1985}. It is however, in better accord
with the very short orbital period of the binary and our
rather rough estimate of K2/5V.

\subsection{V2610~Oph}

V2610~Oph (HD~162905, SAO~141948) was found to be 
variable by \citet{wils2003}
using the Stardial images. The authors 
classified it as a W~UMa-type binary with 
an amplitude of about 0.16 mag and noted that its 
spectral type, K0, is too late for the observed 
orbital period. Later \citet{tas2006} obtained 
precise $BVR$ light curves and determined the 
preliminary geometric parameters of
the system as $q = 0.55$, $i = 54.22\degr$ and found that the system is 
detached but close to contact. The authors 
assumed no third light in the system.
No spectroscopic observations of the system have been published yet.

Our spectroscopy immediately revealed a rather 
different picture: V2610~Oph is
a quadruple system. It is somewhat similar 
to VW~LMi \citep{ddo11} as
it consists of an eclipsing contact binary 
and a detached non-eclipsing pair.
The orbit of the second binary, with a 
period of $P = 8.47$ days, is slightly eccentric 
(see Table~\ref{tab4} and Fig.~\ref{fig5}).

The multiplicity of V2610~Oph makes the
photometric solution of \citet{tas2006} completely inapplicable.
The spectroscopic mass ratio, $q = 0.289$ and the large projected
masses of the components, $(M_1 + M_2) \sin^3 i = 1.408 M_\odot$, 
indicate a much higher inclination angle than what was found previously. 
The light contribution of the second binary
in V2610~Oph around the maxima of the eclipsing pair, 
as found by modeling the
BFs, is $(L_3 + L_4)/(L_1 + L_2) = 1.28 \pm 0.11$, 
i.e., the non-eclipsing pair is brighter than the contact binary. 
When we correct the observed photometric amplitude, 
$\Delta V = 0.163$ mag \citep{tas2006} for this contribution, 
we obtain the full amplitude of the binary as $0.41 \pm 0.02$ mag, 
indicating a high orbital inclination angle.

The center-of-mass RV of the second pair, 
$V_0 = 60.76 \pm 0.11$ km~s$^{-1}$ is fairly close
 to the systemic velocity of the eclipsing 
binary, $71.70 \pm 1.27$ km~s$^{-1}$. 
The RV difference indicates mutual orbital 
motion of the binaries. No systemic-velocity 
changes in either of the binaries were 
noted during our observing run that extended for 
nearly one year. Hence, the outer orbital period is at least 
several years long. The angular separation between the binaries must
be smaller than about 1 arcsec, because V2610~Oph appeared to 
be a single star on our spectrograph slit even during the nights 
of excellent seeing; besides, the system is not known to be a visual
binary \citep{wds}.

A new LC analysis using the new spectroscopic 
mass ratio and including the third light is necessary 
to determine the inclination angle. Then, having the masses of
the close binary, masses of all four components could be determined
using the ``linked mass-ratio'' technique 
as used in \citet{ddo11} for VW~LMi. 

The infrared color of the system, $J-K = 0.393$, 
corresponds to G4V spectral
type, while the TYCHO color, $B-V = 0.587$ 
corresponds to G0V spectral type. 
A direct spectral classification is difficult to perform
but gives a reasonable range of admissible types is F8 -- G2V
which is consistent with the projected mass of the 
dominant primary component of the 
contact binary, $M_1 = 1.09 M_\odot$. 
V2610~Oph was not observed by the Hipparcos satellite and its
trigonometric parallax is unknown.

\subsection{V1387~Ori}

V1387~Ori (HD~42969, HIP~29186) is a member of a trapezoidal, visual, 
multiple system. The other members are GSC~1318~59 and GSC~1318~317; 
the latter star forms the close visual binary HDS~838. 
Neither of these visual 
companions entered our spectrograph slit. Establishing the physical bond 
between the members of this mini-cluster 
would require precise proper motions and RVs.

The variability of V1387~Ori was discovered during 
the Hipparcos mission \citep{hip}.
In the Hipparcos catalogue it was classified as a $\beta$~Lyrae system, but 
later \citet{duer1997} suggested it to be a contact, W UMa-type system. The
Hipparcos light curve shows a maximum following 
the primary minimum that is substantially
brighter than the other maximum 
(similar to the contact binaries AG~Vir and DU~Boo).
No ground-based photometric nor spectroscopic observations of the system
have been published yet.

Our spectroscopy revealed that V1387~Ori 
is a triple system. The radial velocity of
the third component appears to vary as indicated
by our first spectra obtained at HJD 2454073.7873, 
giving $RV_3 = 34.43$ km~s$^{-1}$. This observation was well  
separated in time from the remaining observations which give a 
constant velocity of $RV_3 = 24.4 \pm 0.4$ km~s$^{-1}$. 
The contribution of third component is affected by
the large O'Connell effect observed in the system: 
at Max~I, $L_3/(L_1 + L_2) = 0.128 \pm 0.012$ while at 
Max~II, $L_3/(L_1 + L_2) = 0.186 \pm 0.023$. 
When the third-body signatures are removed from the 
BFs, the orbit of the eclipsing pair is well defined with a low mass ratio,
$q = 0.165 \pm 0.005$, which would be rather
typical for an A-type contact binary. 
No surface inhomogeneities on either of components were noticed 
in the BFs in spite of the large O'Connell effect.

The Hipparcos parallax for this system is negative, 
$\pi = -7.86 \pm 1.96$ mas, 
which most probably results from V1387~Ori 
being located in the visual multiple system.
The 2MASS color, $J-K = 0.155$, corresponding to A2V is  
consistent with our spectral classification, A4V.

\subsection{AU~Ser}

AU~Ser (GSC~1502~1762) is a rather faint ($V_{max} = 10.9$) contact binary 
discovered by \citet{hoff1935}. The first photoelectric photometry presented 
by \citet{binn1972} showed a significant asymmetry in the LC 
(Max~I - Max~II = $-0.05$ mag) and a 0.2 mag difference in the depth
of the eclipses. The light curves obtained 
by \citet{binn1972} were later analyzed
by \citet{kalu1986} who interpreted the large asymmetry as a hot spot
located close to the ``neck'' connecting the components.
The best solution, under the assumption of 
the A-type of the light curve,
was obtained for a rather large mass ratio $q = 0.80$. 
The authors, however, admitted that the acceptable range for the 
mass ratio is rather wide, between 0.70 and 1.15. 
Later \citet{guro2005} analyzed 
photometric observations of this target covering the 
period from 1969 until 2003. Their analysis 
of the observed times of minima indicates the presence of a 
third body on a 94.15 year orbit with an estimated 
minimum mass of $\sim$$M_3 = 0.53 M_\odot$.
According to the authors the differences in 
the LC maxima levels appear to be cyclic on a time scale of about 
30 years.
 
Spectroscopic observations of this system 
were performed by \citet{hriv1993}, 
who reported $q = 0.71$ and a projected total mass of the system of 
$(M_1 + M_2) \sin^3 i = 1.51 M_\odot$. The systemic velocity of the system
was not given. Our new spectroscopic observation leads 
to a projected mass and a mass ratio 
within the errors of those found by \citet{hriv1993}. 
We do not see evidence for a third component 
in the BFs. The phased BFs, however, show a large cool spot on the secondary 
component which is visible after the second 
quadrature (see Fig.~\ref{fig4}). This dark, localized spot 
significantly deforms the BFs and affects the determined RVs. A reliable 
determination of the spectroscopic elements would require modeling of 
simultaneous spectroscopic and photometric observations. 

The system appears to be of relatively late spectral type, 
G4V, for its orbital period of 0.386 d.     
AU~Ser was too faint for the Hipparcos mission, 
hence its parallax is unknown.

\subsection{FT~UMa}

FT~UMa (HD~75840, HIP~43738) was discovered to be variable by the 
Hipparcos satellite.
In the special General Catalogue of Variable Stars 
namelist \citep{namelist74}, the RRc classification
was suggested with a pulsational period of 0.3273519 days. 
High precision $BVR$ photometric observations
of FT~UMa were published by \citet{selam2007}. 
The authors correctly classified the system as a close binary, 
obtained a low mass ratio of $q = 0.25 \pm 0.01$, a
low orbital inclination of $i = 54.48 \pm 0.80\degr$, and suggested
a contact configuration for the system. In spite of 
the fact that this target is relatively bright, $V_{max} = 9.25$, 
and the photometric amplitude is as large 
as 0.17 mag, no other observations of FT~UMa have been published yet.

Our spectroscopy clearly shows that FT~UMa i
s a triple system containing a
close eclipsing binary with a period of 
$P = 0.6547038$ days. FT~UMa is not listed 
in the WDS catalogue as a visual binary. 
During our spectroscopic observations the
star always appeared as a single object, which sets an 
upper limit to the angular separation of the components at about 1 arcsec. 
The center-of-mass velocity of the close binary, $V_0 = -33.66 \pm 1.30$ 
km~s$^{-1}$, is significantly different from the third-component velocity 
which slowly changed from $-22$ km~s$^{-1}$ to about $-10$ km~s$^{-1}$ 
during our run. Unfortunately, the RVs of the 
close pair are rather imprecise to resolve 
the hierarchy and multiplicity of the 
system: it can either be a hierarchical triple 
or a quadruple system consisting of 
two binaries (with the non-eclipsing pair being the SB1). 
The light contribution of the third component around the maxima 
of the eclipsing pair is $L_3/(L_1+L_2) = 1.01 \pm 0.03$.
The full, undiluted, photometric amplitude of the eclipsing binary would be 
then about 0.37 mag. The presence of such a 
strong third light made the photometric
solution of \citep{selam2007} of limited use.

The profiles of all three components are well separated in the extracted BFs, 
indicating that the close eclipsing binary is very probably detached, which 
is consistent with the large mass ratio, $q = M_2/M_1 = 0.984 \pm 0.019$, 
but rather atypical for contact binaries. On the other hand, 
Hipparcos photometry phased with the adopted period of 0.6547038 days 
results in a LC that is typical of contact binaries. 

The TYCHO2 color 
of the system, $B-V = 0.404$ reflects the combined contributions
of the eclipsing pair and of the third component. 
Our spectral classification spectra suggest the F0V spectral type, which
is consistent with the 2MASS $J-K$ = 0.207.
The Hipparcos parallax is fairly small, $6.90 \pm 1.28$ mas,
and too inaccurate to be of use.

\section{SUMMARY}
\label{summary}

With the ten new short-period binaries, this paper brings the
number of the systems studied at the David Dunlap Observatory to
one-hundred thirty. With the closure of the observatory,
this series is coming to an end. However, the number
of known, bright close binaries that remained 
unobserved at DDO is moderate; only a dozen or so 
known W~UMa-type eclipsing binaries 
brighter than about $V = 10$ that were accessible from DDO 
remained. We plan to
publish the data for the unfinished cases in the last, 
fifteenth installment of this series. 

The highlights of the current series are: 
(1)~the discoveries of two quadruple systems 
TZ~Boo and V2610~Oph, (2)~the discoveries of four triple systems: 
EL~Boo, GK~Cep, V1387~Ori, and FT~UMa, (3)~the spotted contact 
binary AU~Ser, with a large spot on the secondary component.
None of the detected spectroscopic multiple systems were 
previously noted to be visual binaries (see \citet{wds}).
 
While for four systems, EL~Boo, V2610~Oph, V1387~Ori and FT~UMa, we are
presenting the first spectroscopic observations, the quality of the data
for the remaining six systems is much improved relative to the previously
published investigations.

Numerous discoveries and reliable
solutions of triple and quadruple systems show that the
BF deconvolution approach utilizing the SVD method is a 
powerful technique. In this series, good examples 
are the three triple systems with a dominant third component,
EL~Boo, V2610~Oph and FT~UMa. Also, the BF technique enabled us to 
determine the first reliable set of 
spectroscopic elements for TZ~Boo and to reveal
the true nature of this system. The third component of this system, 
moving in 9.48 day orbit, blends with the components of the 
close binary, a condition that made all previous spectroscopic studies
of TZ~Boo unsuccessful. 
The second binary may be the cause of the large perturbations
observed in the light curve of the eclipsing pair. However, 
with several multiple systems analyzed in the DDO papers,
we also see the weakness of the BF technique that should
be addressed in the future: As described in \citet{rucpri2008}, the  
contribution of the third component to the total light of the system
is always overestimated. There are two reasons for this
discrepancy: (i)~the continuum spectrum
rectification differently affects spectra of 
(usually) sharp-lined companions and of heavily broadened
short-period binaries; (ii)~the technique must use a 
single template spectrum, but this affects the luminosity ratio
when the spectral types
of the third component and the contact binary are significantly 
different.

\acknowledgements

A grant to SMR and the graduate scholarship to BC 
from the Natural Sciences and Engineering Council of Canada 
are acknowledged with gratitude. 
Thanks are extended to the Polish Science Committee for
grants PO3D~006~22, P03D~003~24, N20300731 and N203304335.
The travel of TP to 
Canada has been supported by a Slovak Academy of 
Sciences VEGA grant 2/7010/7. In 2008, TP has been a 
recipient of the Post-Doctoral Fellowship of the
Canadian Space Agency; he appreciates the hospitality and 
support of the local staff during his stay at DDO.

The research made use of the SIMBAD database, operated at the CDS,
Strasbourg, France and accessible through the Canadian
Astronomy Data Centre, which is operated by the Herzberg Institute of
Astrophysics, National Research Council of Canada.
This research made also use of the Washington Double Star (WDS)
Catalog maintained at the U.S. Naval Observatory.

\clearpage

\noindent
Captions to figures:

\bigskip

\figcaption[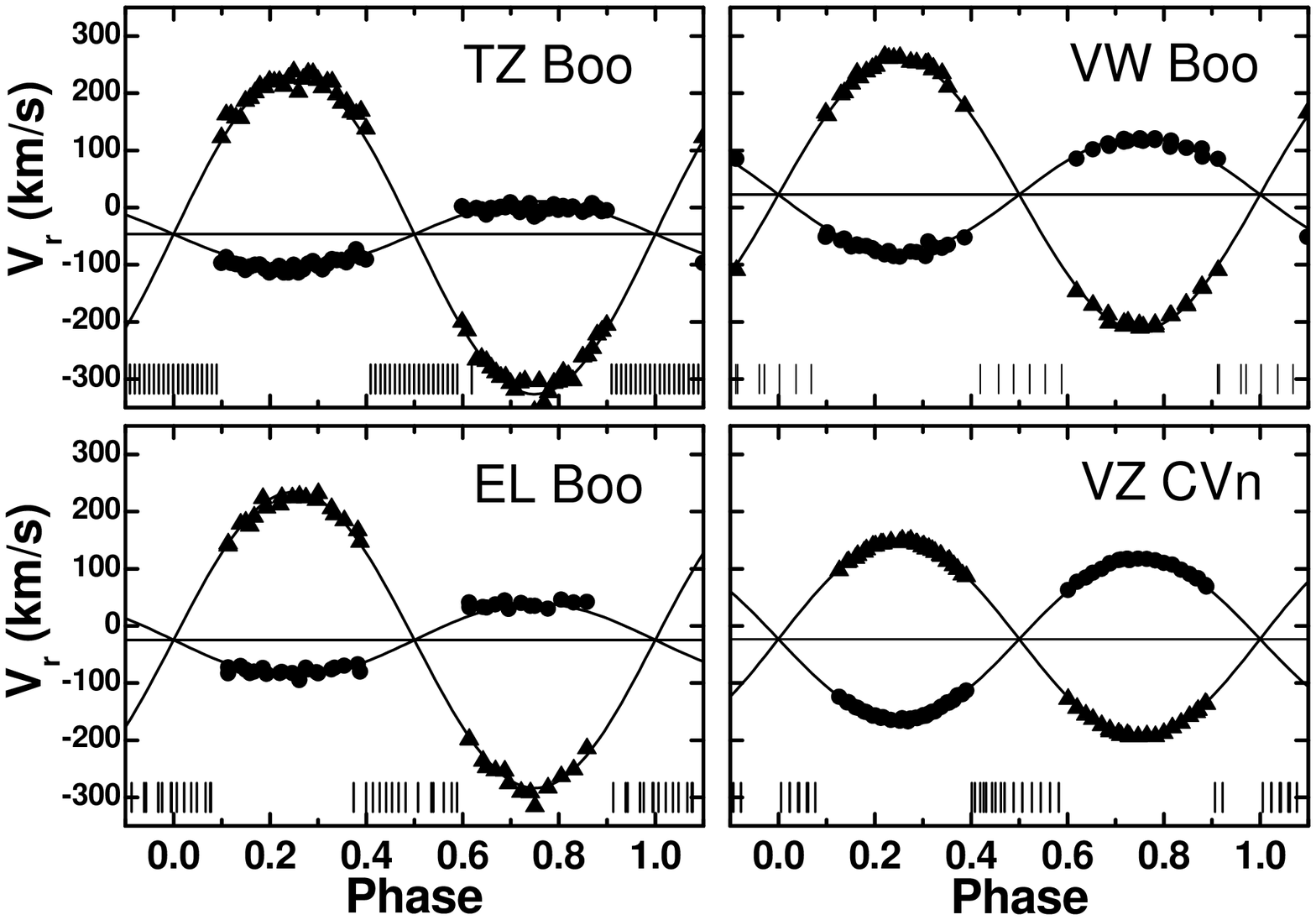] {\label{fig1} 
Radial velocities of the
systems TZ~Boo, VW~Boo, EL~Boo, and VZ~CVn are plotted in individual 
panels versus the orbital phases. The lines give the respective 
circular-orbit (sine-curve) fits to the RVs.
TZ~Boo is a quadruple system consisting of a contact and 
a detached single-line
binary while VW~Boo is a contact binary. 
EL~Boo is a triple system harboring
a close binary and VZ~CVn is a detached or a semi-detached binary.
The circles and triangles in this and the next two figures correspond 
to components with velocities $V_1$ and $V_2$, as listed in Table~\ref{tab1}, 
respectively. The component eclipsed at the minimum corresponding to 
$T_0$ (as given in Table~\ref{tab2}) is the one that 
shows negative velocities for
the phase interval $0.0 - 0.5$ and is the more massive one.
Short marks in the lower parts of the panels show phases of available
observations which were not used in the solutions because of the
spectral line blending. 
}

\figcaption[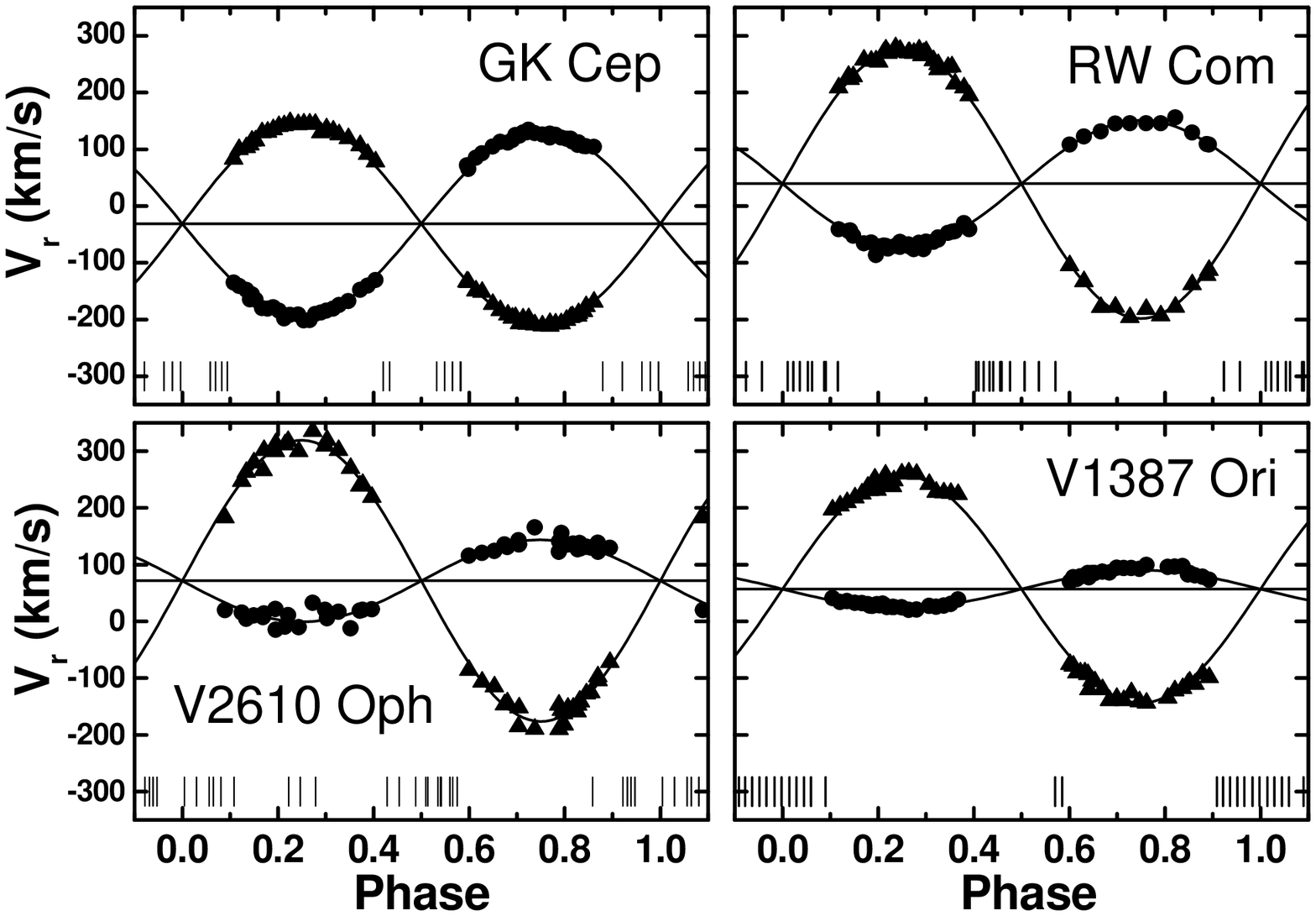] {\label{fig2} 
The same as for
Figure~\ref{fig1}, but for GK~Cep, RW~Com, V2610~Oph, and V1387~Ori.
GK~Cep is a close binary with a faint third component; RW~Com is a contact 
binary; V2610~Oph is a quadruple system consisting of a contact eclipsing 
binary and a wide non-eclipsing pair; V1387~Ori
is a triple system containing a close eclipsing binary. V1387~Ori forms a
part of a trapezoidal visual multiple system. 
}

\figcaption[rvs9-10.ps] {\label{fig3} 
The same as for
Figures~\ref{fig1} and \ref{fig2}, for the two remaining systems
AU~Ser, and FT~UMa. AU~Ser is a contact binary while 
FT~UMa is a close eclipsing binary in a triple system. 
}

\figcaption[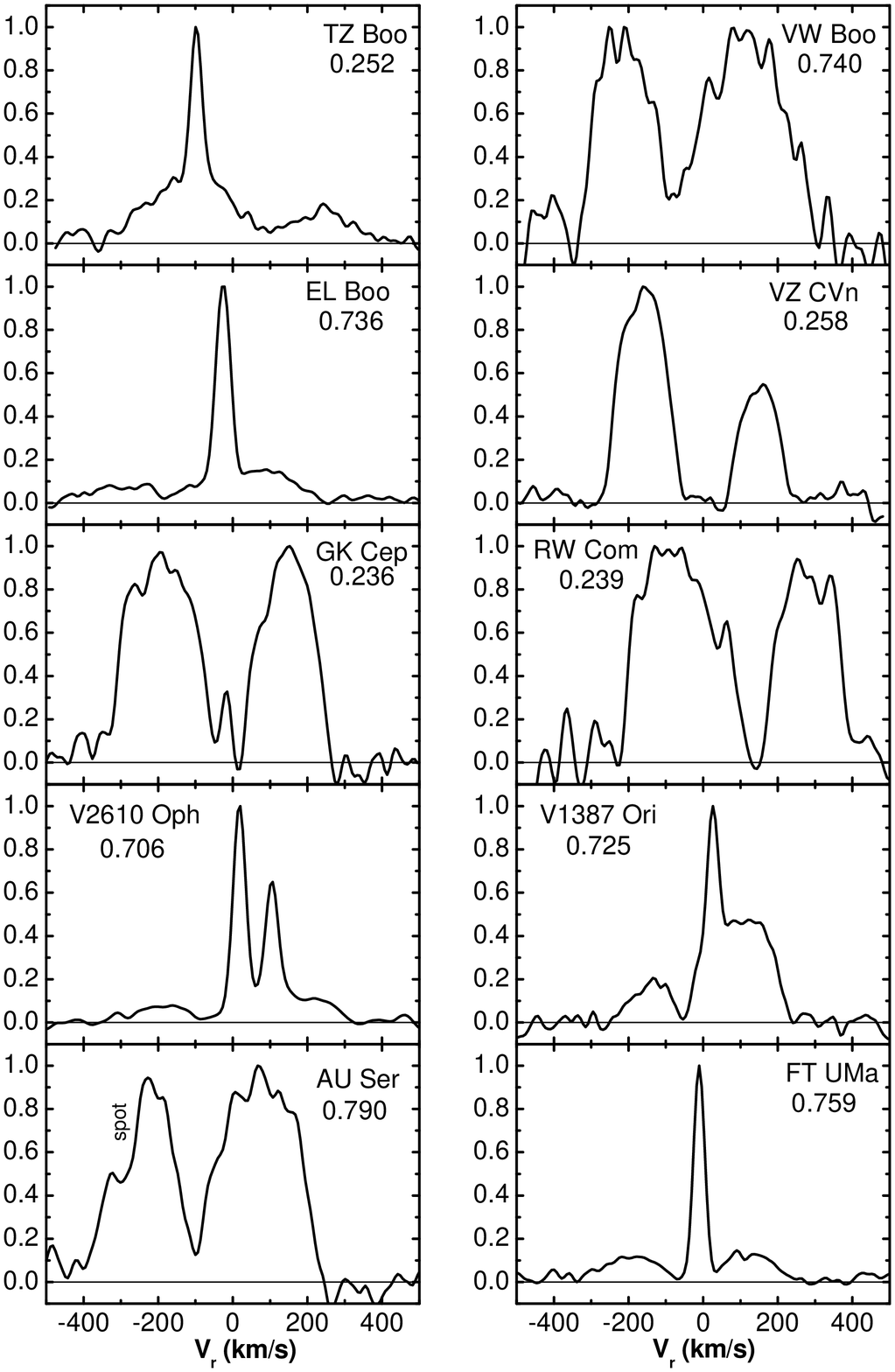] {\label{fig4}
The broadening functions (BFs) for all ten systems of 
this group, selected for orbital phases close to 0.25 or 0.75. 
The phases are marked by numbers in the
individual panels. Additional components to the close binaries, TZ~Boo,
EL~Boo, V2610~Oph, V1387~Ori, and FT~UMa, are strong and clearly visible.
The third component in GK~Cep is a small peak close to the 
binary center-of-mass velocity.
All panels have the same horizontal range, $-500$ to 
$+500$ km~s$^{-1}$. 
}

\figcaption[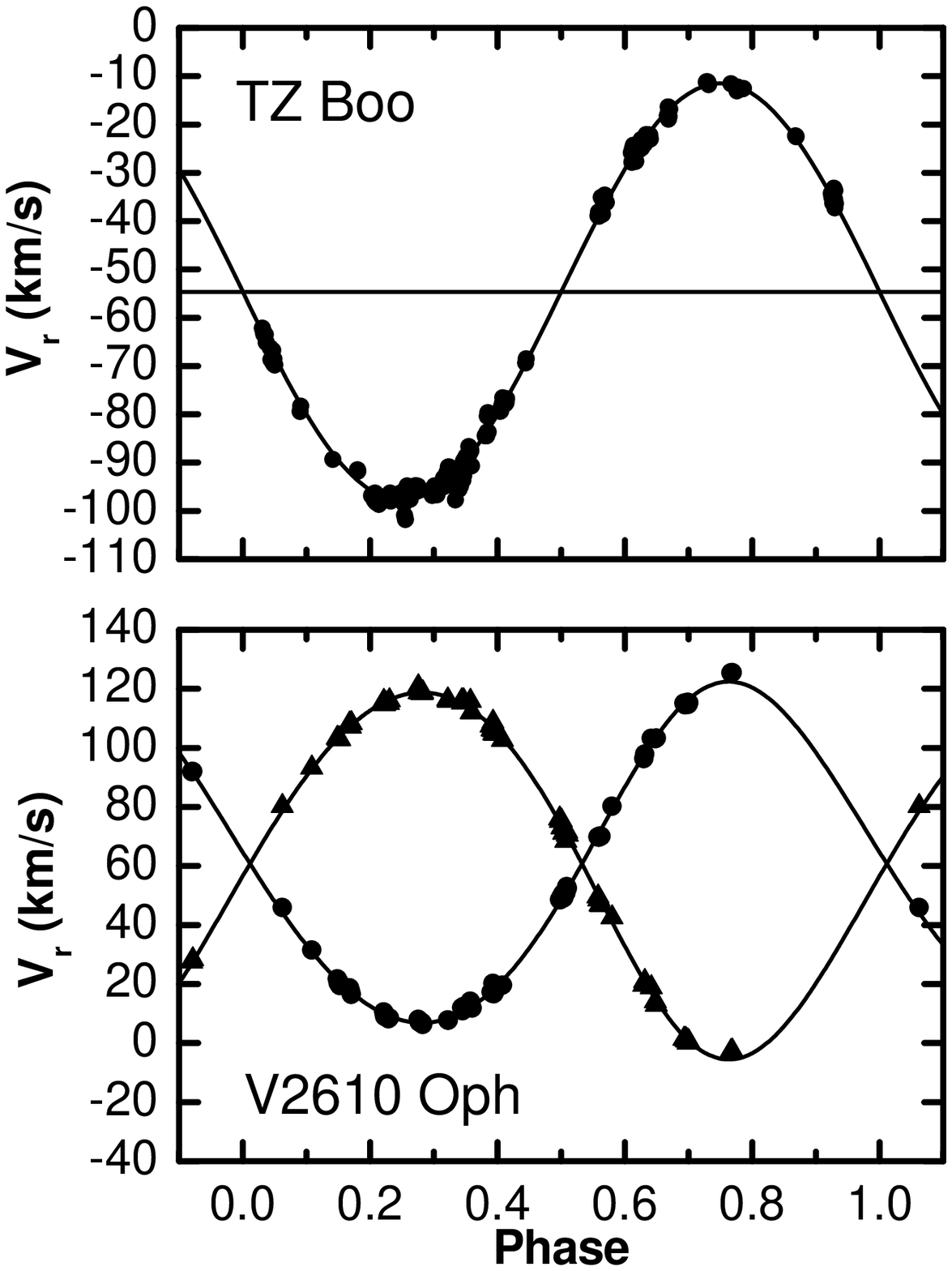] {\label{fig5}
The radial velocities and the best fits 
for the third component of TZ~Boo and 
the second non-eclipsing binary in the quadruple system
V2610~Oph.}

\clearpage

\addtocounter{figure}{-5}

\begin{figure} 
\epsscale{0.85}
\plotone{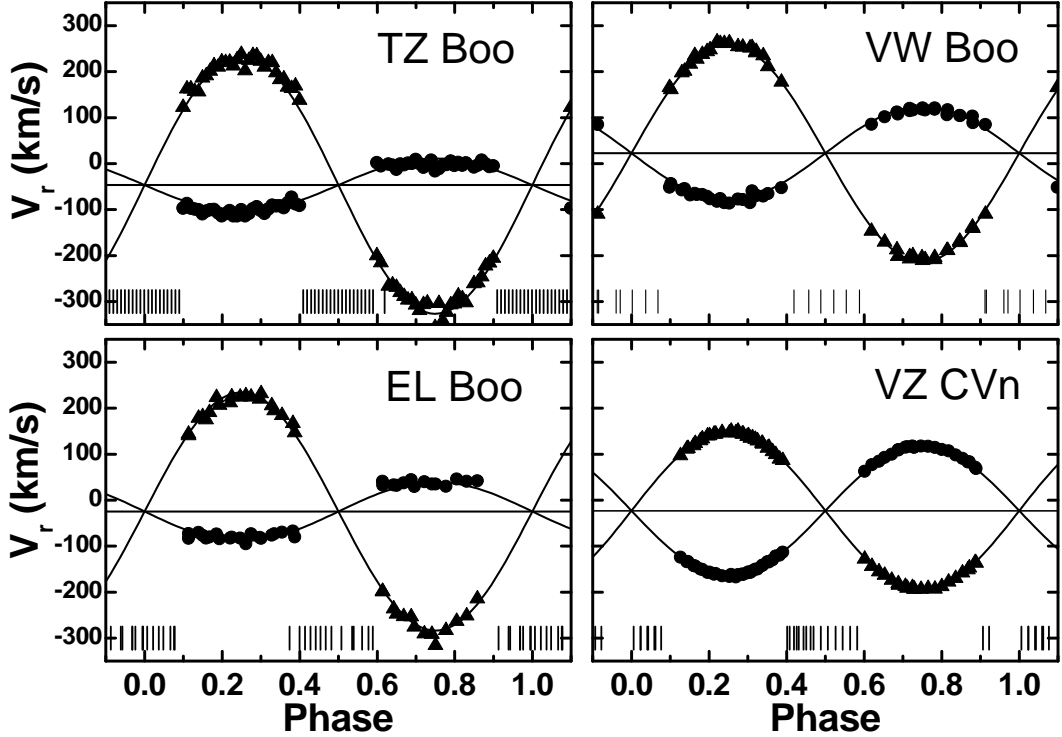}
\caption{Radial velocities of the
systems TZ~Boo, VW~Boo, EL~Boo, and VZ~CVn are plotted in individual 
panels versus the orbital phases. The lines give the respective 
circular-orbit (sine-curve) fits to the RVs.
TZ~Boo is a quadruple system consisting of a contact and 
a detached single-line
binary while VW~Boo is a contact binary. 
EL~Boo is a triple system harboring
a close binary and VZ~CVn is a detached or a semi-detached binary.
The circles and triangles in this and the next two figures correspond 
to components with velocities $V_1$ and $V_2$, as listed in Table~\ref{tab1}, 
respectively. The component eclipsed at the minimum corresponding to 
$T_0$ (as given in Table~\ref{tab2}) is the one that 
shows negative velocities for
the phase interval $0.0 - 0.5$ and is the more massive one.
Short marks in the lower parts of the panels show phases of available
observations which were not used in the solutions because of the
spectral line blending. 
}
\end{figure}

\begin{figure} 
\epsscale{0.85}
\plotone{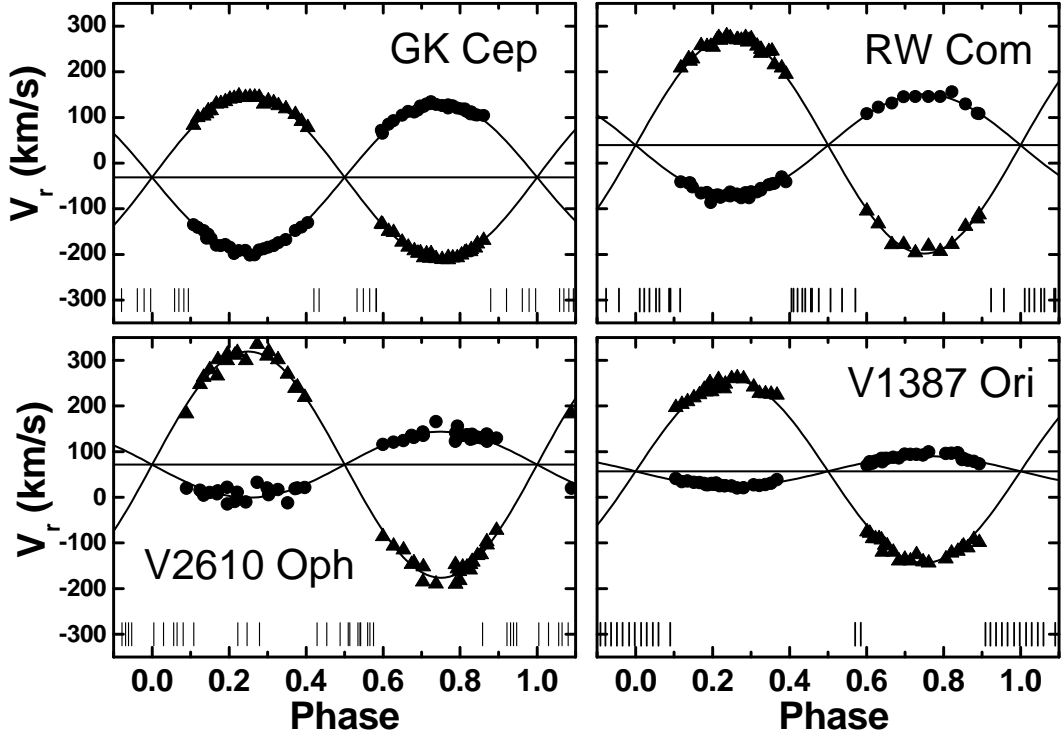}
\caption{The same as for
Figure~\ref{fig1}, but for GK~Cep, RW~Com, V2610~Oph, and V1387~Ori.
GK~Cep is a close binary with a faint third component; RW~Com is a contact 
binary; V2610~Oph is a quadruple system consisting of a contact eclipsing 
binary and a wide non-eclipsing pair; V1387~Ori
is a triple system containing a close eclipsing binary. V1387~Ori forms a
part of a trapezoidal visual multiple system. 
}
\end{figure}

\begin{figure} 
\epsscale{0.85}
\plotone{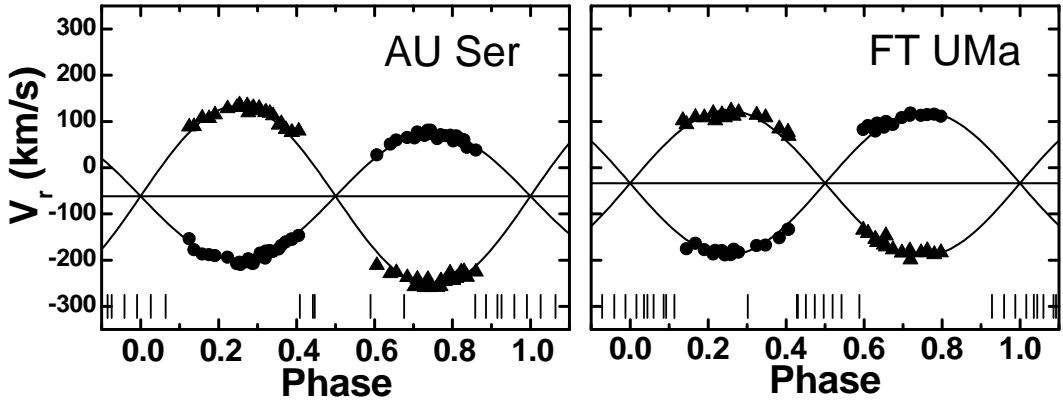}
\caption{The same as for
Figures~\ref{fig1} and \ref{fig2}, for the two remaining systems
AU~Ser, and FT~UMa. AU~Ser is a contact binary while 
FT~UMa is a close eclipsing binary in a triple system. 
}
\end{figure}

\begin{figure} 
\epsscale{0.75}
\plotone{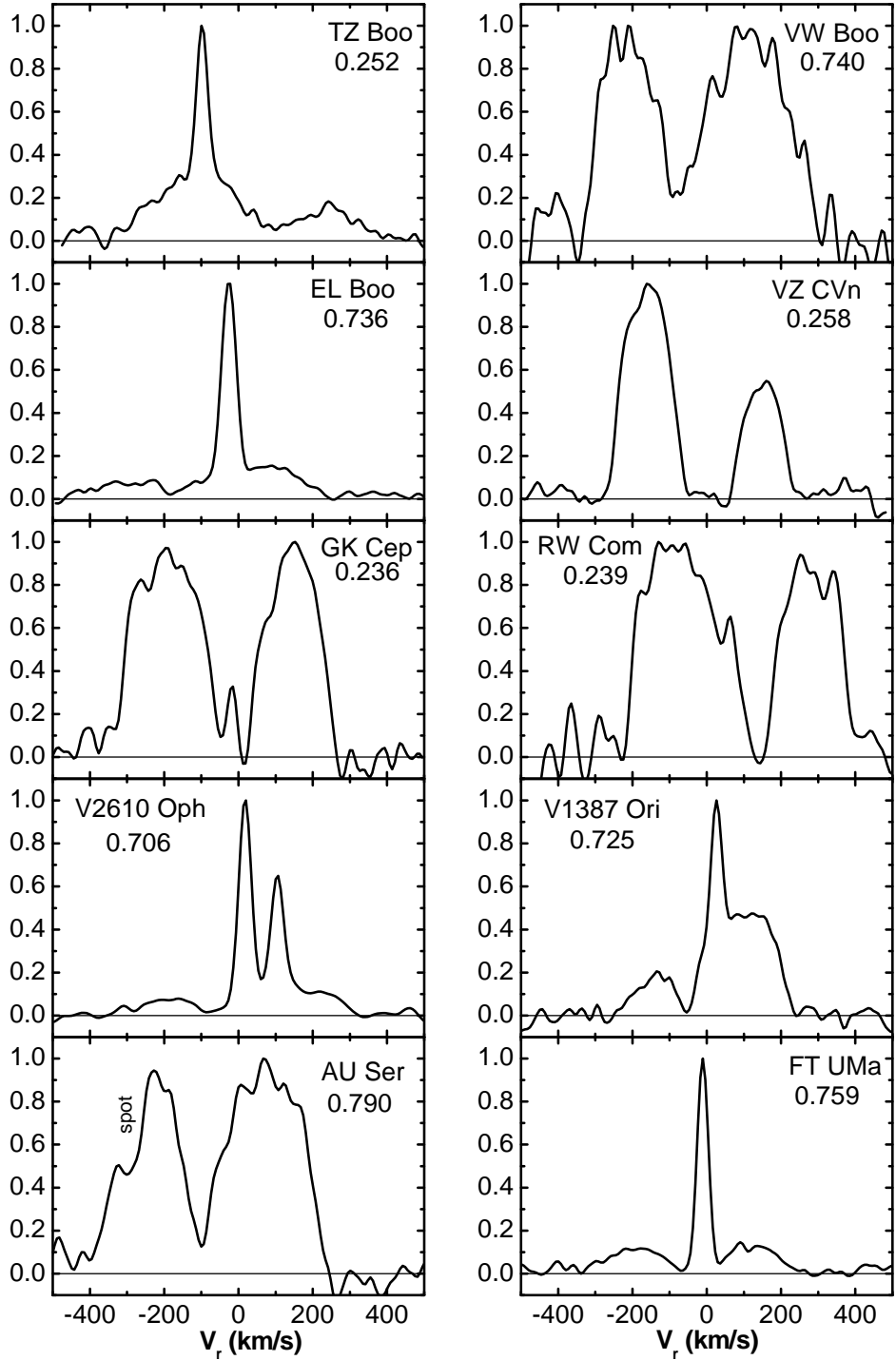}
\caption{The broadening functions (BFs) for all ten systems of 
this group, selected for orbital phases close to 0.25 or 0.75. 
The phases are marked by numbers in the
individual panels. Additional components to the close binaries, TZ~Boo,
EL~Boo, V2610~Oph, V1387~Ori, and FT~UMa, are strong and clearly visible.
The third component in GK~Cep is a small peak close to the 
binary center-of-mass velocity.
All panels have the same horizontal range, $-500$ to 
$+500$ km~s$^{-1}$. 
}
\end{figure}

\begin{figure} 
\epsscale{0.75}
\plotone{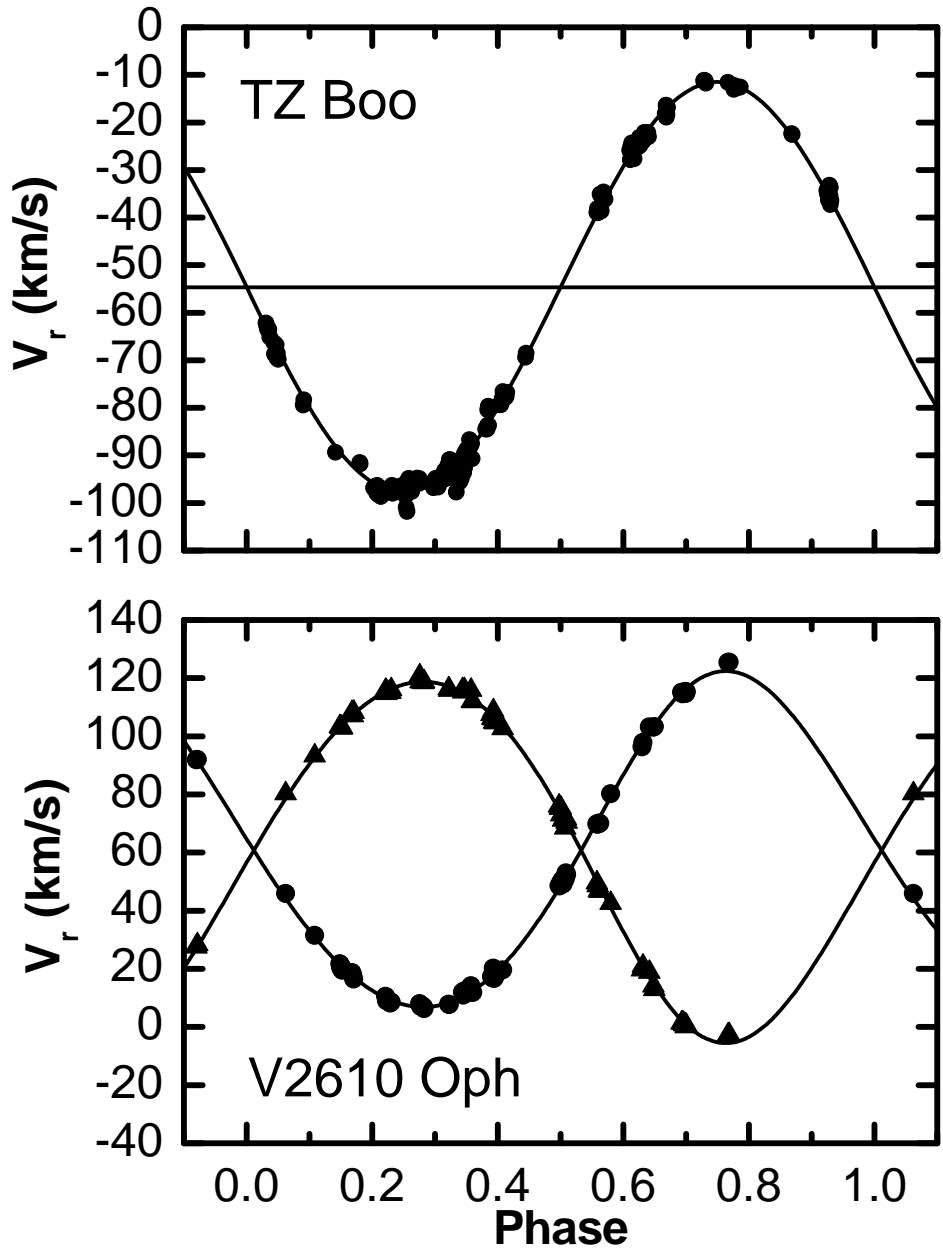}
\caption{The radial velocities and the best fits 
for the third component of TZ~Boo and 
the second non-eclipsing binary in the quadruple system
V2610~Oph.
}
\end{figure}

\begin{deluxetable}{llrrrrr}

\tabletypesize{\footnotesize}

\tablewidth{0pt}
\tablenum{1}

\tablecaption{DDO radial velocity observations (the full
table is available only in the electronic form)
\label{tab1}}

\tablehead{
\colhead{Target}          &
\colhead{HJD--2,400,000}  &
\colhead{~$V_1$}          &
\colhead{~~$W_1$}         &
\colhead{~$V_2$}          &
\colhead{~~$W_2$}         &
\colhead{Phase}          \\
                          &
                          &
\colhead{[km s$^{-1}$]}   &
                          &
\colhead{[km s$^{-1}$]}   &
                          & \\
}
\startdata
VW~Boo & 54558.8318 &   0.00  &  0.00  &    0.00 &  0.00  & 0.9153 \\
VW~Boo & 54558.8472 &   0.00  &  0.00  &    0.00 &  0.00  & 0.9602 \\
VW~Boo & 54558.9127 &$-68.22$ &  0.50  &  216.48 &  0.50  & 0.1516 \\
VW~Boo & 54558.9233 &$-67.24$ &  1.00  &  237.55 &  1.00  & 0.1826 \\
VW~Boo & 54559.7940 & 116.90  &  1.00  &$-200.04$&  1.00  & 0.7261 \\
VW~Boo & 54559.8047 & 116.91  &  1.00  &$-208.79$&  1.00  & 0.7574 \\
VW~Boo & 54563.6869 &$-51.23$ &  1.00  &  165.41 &  1.00  & 0.0984 \\
VW~Boo & 54563.6977 &$-57.31$ &  1.00  &  197.76 &  1.00  & 0.1299 \\
VW~Boo & 54563.7091 &$-64.92$ &  1.00  &  235.53 &  1.00  & 0.1632 \\
VW~Boo & 54563.7200 &$-71.21$ &  1.00  &  243.67 &  1.00  & 0.1951 \\
\enddata

\tablecomments{The table gives the RVs $V_i$ for observations 
described in the paper. The first 10 rows of the table are shown for one 
of our typical program stars, VW~Boo. For first program star, TZ~Boo, where
phase smoothed BFs were used, the heliocentric Julian dates are not given.
Observations leading to entirely inseparable broadening function peaks 
are given zero weight; these observations may be eventually used in more 
extensive modeling of the broadening functions. Zero weights were assigned to 
observations of marginally visible peaks of the secondary (sometimes
even the primary) component. The RVs designated as $V_1$ correspond to
the more massive component; it was always the component eclipsed
during the minimum at the epoch $T_0$ (this does not always correspond
to the deeper minimum and photometric phase 0.0). The phases
correspond to spectroscopic $T_0$ and periods given in Table~2, but not
necessarily to the photometric ephemerides given below the table). }

\end{deluxetable}

\begin{deluxetable}{lccrrrccc}

\tabletypesize{\scriptsize}

\pagestyle{empty}
\tablecolumns{9}

\tablewidth{0pt}

\tablenum{2}
\tablecaption{Spectroscopic orbital elements \label{tab2}}
\tablehead{
   \colhead{Name} &                
   \colhead{Type} &                
   \colhead{Other names} &         
   \colhead{$V_0$~~~} &            
   \colhead{$K_1$~~~} &            
   \colhead{$\epsilon_1$~} &       
   \colhead{T$_0$ -- 2,400,000} &  
   \colhead{P (days)} &            
   \colhead{$q$}          \\       
   \colhead{}     &                
   \colhead{Sp.~type}    &         
   \colhead{}      &               
   \colhead{} &                    
   \colhead{$K_2$~~~} &            
   \colhead{$\epsilon_2$~} &       
   \colhead{$(O-C)$(d)~[E]} &      
   \colhead{$(M_1+M_2) \sin^3i$} & 
   \colhead{}                      
}
\startdata
TZ~Boo     & EW(A/W)             & BD+40 2857  & $-46.57$(0.90)   & 57.8(1.45)  &
      8.70 & 54042.7482(4)       & 0.2971597   & 0.207(5)         \\ 
           & F/G5                & HIP~74061   &                  &280.02(1.50) &
     10.49 & $+0.0002$~[+5.191.0]& 1.188(18)   &                  \\[1mm]
VW~Boo     & EW                  &             & $22.79$(0.66)    & 101.39(1.07)&
      6.65 & 54573.9227(4)       & 0.3423157   & 0.428(5)         \\ 
           & G5V                 & HIP~69826   &                  & 236.74(1.03)&
      5.99 & $-0.0032$~[+6,058.5]& 1.371(14)   &                  \\[1mm]
EL~Boo     & EW                  & BD+14 2788  & $-24.58$(1.12)   & 64.19(1.70) &
      9.58 & 54584.1779(9)       & 0.413772    & 0.248(7)         \\ 
           & F5V                 & HIP~72391   &                  &259.07(1.89) &
      9.31 & $-0.0850$~[+14,704] & 1.448(28)   &                  \\[1mm]
VZ~CVn     & EB                  & HD~117777   & $-22.99$(0.16)   &141.51(0.26) &
      1.64 & 54573.9540(3)       & 0.84246123  & 0.8252(19)       \\ 
           & F0V                 & HIP~66017   &                  &171.48(0.25) &
      1.72 & $+0.0009$~[+2,461]  & 2.676(7)    &                  \\[1mm]
GK~Cep     & EB                  & HD~205372   & $-31.06$(0.45)   & 163.44(0.73)&
      5.05 & 54464.4014(8)       & 0.936169    & 0.913(5)         \\ 
           & A0V                 & HIP~106226  &                  & 178.95(0.73)&
      4.20 & $-0.0062$~[+2,098]  & 3.893(26)   &                  \\[1mm]
RW~Com     & EW(W)               &             & $39.35$(0.86)    & 112.04(1.28)&
      6.33 & 54272.7645(3)       & 0.2373464(7)& 0.471(6)         \\ 
           & K2/5V               & HIP~61243   &                  & 237.70(1.29)&
      7.39 & $+0.0005$~[+7,465.5]& 1.052(13)   &                  \\[1mm]
V2610~Oph  & EW(A)               & HD~162905   & $71.70$(1.27)    & 72.06(2.06) &
     11.87 & 54461.5659(10)      & 0.426512(3) & 0.291(9)         \\ 
           & F8/G2V              &             &                  &247.42(2.08) &
     12.11 & $+0.0109$~[+4,904]  & 1.441(31)   &                  \\[1mm]
V1387~Ori  & EW?                 & HD~42069    & $56.92$(0.65)    & 33.21(0.98) &
      4.37 & 54204.0187(13)      & 0.730166    & 0.165(5)         \\ 
           & A4V                 & HIP~29186   &                  &200.72(1.18) &
      8.50 & $+0.0082$~[+7,811]  & 0.969(16)   &                  \\[1mm]
AU~Ser     & EW(A)               &             & $-61.59$(0.84)   &138.77(1.33) &
      6.36 &  54597.4761(7)      & 0.386499    & 0.709(8)         \\ 
           & G4V                 &             &                  &195.64(1.34) &
     10.43 & $+0.0042$~[+5,426]  & 1.498(19)   &                  \\[1mm]
FT~UMa     & EB                  & HD~75840    & $-33.66$(1.30)   &155.15(2.14) &
     10.78 & 54561.9614(19)      & 0.654704    & 0.984(19)        \\ 
           & F0V                 & HIP~43738   &                  &157.73(2.13) &
     10.88 & $-0.0136$~[+9,258.5]& 2.077(45)   &                  \\[1mm]
\enddata
\tablecomments{The spectral types given in column 2 
relate to the combined spectral
type of all components in a system; they are given 
in parentheses if taken from the
literature, otherwise they are new. The convention of naming 
the binary components in the  table is that the 
more massive star is marked by the subscript ``1'', so that the
mass ratio is defined to be always $q \le 1$. 
The standard errors of the circular solutions in 
the table are expressed in units of the last decimal 
places quoted; they  are given in parentheses 
after each value. The center-of-mass velocities ($V_0$),
the velocity amplitudes ($K_i$) and the standard 
unit-weight errors of the solutions
($\epsilon$) are all expressed in km~s$^{-1}$. 
The spectroscopically determined  moments of primary 
or secondary minima are given by $T_0$ (corresponding 
approximately to the average Julian date of the run); 
the corresponding $(O-C)$ deviations (in days) have been 
calculated from the available prediction on
primary minimum, as given in the text, using 
the assumed periods and the number of
epochs given by [E]. 
The values of $(M_1+M_2)\sin^3 i$ are in the solar mass units.\\
Ephemerides ($HJD_{min}$ -- 2,400,000 + period in days) 
used for the computation of the $(O-C)$ residuals:\\
 TZ~Boo:    52500.1920 + 0.2971597 \\  
 VW~Boo:    52500.0062 + 0.3423157 \\  
 EL~Boo:    48500.1594 + 0.413772  \\  
 VZ~CVn:    52500.6560 + 0.84246123\\  
 GK~Cep:    52500.325  + 0.936169  \\  
 RW~Com:    52500.1432 + 0.2373463 \\  
 V2610~Oph: 52369.95   + 0.42651   \\  
 V1387~Ori: 48500.6839 + 0.730166  \\  
 AU~Ser:    52500.3392 + 0.386497  \\  
 FT~UMa:    48500.3999 + 0.6547038 \\  
}
\end{deluxetable}

\begin{deluxetable}{llrrc}

\tabletypesize{\footnotesize}

\tablewidth{0pt}
\tablenum{3}
\tablecolumns{5}

\tablecaption{Radial velocity observations (the full
table is available only in electronic form) of third and fourth
components of multiple systems \label{tab3}}
\tablehead{
\colhead{Target} & \colhead{HJD--2,400,000} & \colhead{~V$_3$}         & \colhead{~V$_4$}        & \\
                 &                          & \colhead{[km s$^{-1}$]}  & \colhead{[km s$^{-1}$]} & \\
}
\startdata
V2610~Oph & 54302.7053  & 91.91 &  27.54 \\
V2610~Oph & 54302.7160  & 91.99 &  28.35 \\
V2610~Oph & 54306.6868  & 17.32 & 107.57 \\
V2610~Oph & 54306.6976  & 16.85 & 106.22 \\
V2610~Oph & 54306.7090  & 20.25 & 108.84 \\
V2610~Oph & 54306.7197  & 17.23 & 106.02 \\
V2610~Oph & 54307.5974  & 48.53 &  76.07 \\
V2610~Oph & 54307.6080  & 48.49 &  75.25 \\
V2610~Oph & 54307.6193  & 49.66 &  75.37 \\
V2610~Oph & 54307.6299  & 49.44 &  73.04 \\
\enddata

\tablecomments{The table gives the RVs $V_i$ for
the third and fourth components. The typical 10 rows of the table
for quadruple system, V2610~Oph, are shown. Observations of the quadruple 
system V2610~Oph leading to entirely inseparable broadening function 
peaks of the components of the second binary have been omitted from the table 
and are not used in the computation of the orbit.}
\end{deluxetable}

\begin{deluxetable}{lrr}

\tabletypesize{\footnotesize}

\tablewidth{0pt}
\tablenum{4}

\tablecaption{Spectroscopic orbital elements of the second
non-eclipsing binaries in the quadruple systems TZ~Boo and 
V2610~Oph \label{tab4}. Orbit of the second binary in TZ~Boo
is circular, thus $e_{34} = 0.00$ and $\omega_{34} = \pi/2$}
\tablehead{
\colhead{Parameter} &  & \colhead{error} \\
}
\startdata
\sidehead{\bf TZ~Boo}
$P_{34}$ [days]          & 9.47765  &   0.00034    \\
$T_0$ [HJD]              & 2\,454\,042.818 & 0.007 \\
$V_0$ [km~s$^{-1}$]      & $-$54.64 &   0.12       \\
$K_3$ [km~s$^{-1}$]      &    43.15 &   0.14       \\
$a_3 \sin i$ [R$_\odot$] &  8.08    &   0.03       \\
$f(m)$ [M$_\odot$]       & 0.0793   &   0.0008     \\
\sidehead{\bf V2610~Oph}
$P_{34}$ [days]        & 8.47093  &   0.00025  \\
$e_{34}$               & 0.073    &   0.003    \\
$\omega$ [rad]         & 5.88     &   0.04     \\
$T_0$ [HJD]            & 2\,454\,461.85 & 0.06  \\
$V_0$ [km~s$^{-1}$]    &    60.76 &   0.11     \\
$K_3$ [km~s$^{-1}$]    &    57.77 &   0.23     \\
$K_4$ [km~s$^{-1}$]    &    62.15 &   0.24     \\
$q = K_4/K_3$          &    0.929 &   0.005    \\
$(a_3+a_4)\sin i$ [R$_\odot$] &    20.01 &   0.06     \\
$(M_3+M_4)\sin^3 i$ [M$_\odot$] & 1.502 & 0.013\\
\enddata

\tablecomments{The table gives spectroscopic elements of the second
 binaries in TZ~Boo and V2610~Oph: orbital period ($P_{34}$), eccentricity ($e_{34}$),
 longitude of the periastron passage ($\omega$), time of the periastron
 passage ($T_0$), systemic velocity ($V_0$), semi-amplitudes of the
 RV changes ($K_3,K_4$). Corresponding mass ratio $q$, and projected relative 
 semi-major ($(a_3+a_4)\sin i$) and total mass ($(M_3+M_4)\sin^3 i$) is given 
 for V2610~Oph where both lines of the second binary could be measured. For 
 single-lined non-eclipsing binary in TZ~Boo only $a_3 \sin i$ and $f(m)$ is given.}
\end{deluxetable}

\end{document}